\documentstyle[epsf,aps,12pt]{revtex}

\begin{document}

\def\del{\partial}
\def\Lap{\bigtriangleup}
\def\Tr{{\rm Tr}}
\def\^{\wedge}
\def\goinf{\rightarrow\infty}
\def\goes{\rightarrow}
\def\bm{\boldmath}
\def\-{{-1}}
\def\inv{^{-1}}
\def\sqr{^{1/2}}
\def\isqr{^{-1/2}}

\def\reff#1{(\ref{#1})}
\def\vb#1{{\partial \over \partial #1}} 
\def\Del#1#2{{\partial #1 \over \partial #2}}
\def\Dell#1#2{{\partial^2 #1 \over \partial {#2}^2}}
\def\Dif#1#2{{d #1 \over d #2}}
\def\Lie#1{ {\cal L}_{#1} }
\def\diag#1{{\rm diag}(#1)}
\def\abs#1{\left | #1 \right |}
\def\rcp#1{{1\over #1}}
\def\paren#1{\left( #1 \right)}
\def\brace#1{\left\{ #1 \right\}}
\def\bra#1{\left[ #1 \right]}
\def\angl#1{\left\langle #1 \right\rangle}

\def\wa{&=&}
\def\wb{&\equiv &}

\def\dg#1#2{g_{#1#2}}
\def\ug#1#2{g^{#1#2}}
\def\ux#1{{\dot{x}^{#1}}}
\def\dx#1{{\dot{x}_{#1}}}

\def\KY{Killing-Yano }
\def\FRW{Friedman-Robertson-Walker }
\def\KN{Kerr-Newman }
\def\PF{Penrose-Floyd }
\def\PD{Poisson-Dirac }

\def\chiral{\Gamma_*}
\def\dualQ{Q^*}
\def\Af{{\cal A}_f}
\def\alg{{\cal G}_2}
\def\salg{{\cal H}}

\def\abstract#1{\begin{center}{\large ABSTRACT}\end{center}
\par #1}
\def\title#1{\begin{center}{\large #1}\end{center}}
\def\author#1{\begin{center}{\sc #1}\end{center}}
\def\address#1{\begin{center}{\it #1}\end{center}}

\def\pubnum{277/COSMO-50}

\begin{titlepage}
\hfill
\parbox{6cm}{{TIT/HEP-\pubnum} \par January 1995 }
\par
\vspace{7mm}
\title{The role of Killing-Yano tensors in supersymmetric mechanics
on a curved manifold}
\vskip 1cm
\author{Masayuki TANIMOTO
\footnote{E-mail address: prince@th.phys.titech.ac.jp}}
\address{Department of Physics, Tokyo Institute of \\ Technology,
Oh-Okayama Meguro-ku, Tokyo 152, Japan}
\vskip 1 cm
\abstract{
The supersymmetric extension of charged point particle's motion
is applied to investigate symmetries of gravitational fields and
electromagnetic fields.
We mainly focus on the role of the \KY tensors
of both usual and generalized types.
Results obtained by systematic analysis strengthen
the connection of the \KY tensor and superinvariants
(functions commuting with the supercharge).
}

\vfill
\end{titlepage}

\addtocounter{page}{1}

\section{Introduction}

Recently, Gibbons, Rietdijk and van Holten \cite{GRvH}
investigated symmetries of spacetimes systematically
in terms of the motion of
pseudo-classical spinning point particles described by the
supersymmetric extension of the usual relativistic point particle
\cite{BM,Casa,BCL,BDZVH,BDVH}.
Such a supersymmetric theory possesses a supercharge $Q$ generating
the supersymmetry
transformation between particle's position $x^\mu$ and particle's
``spin'' $\xi^a$, which must be introduced to forbid the negative norm
state of spin due to the indefinite Lorentz metric $\eta_{ab}$.
One outstanding feature of such a theory is to have an algebra
like $\brace{Q,Q}\propto H$, where $H$ is the Hamiltonian.
Due to this relation and the Jacobi identity, superinvariants
$J$ such as $\brace{Q,J}=0$ are simultaneously constants of
motion $\brace{H,J}=0$, so that superinvariants are of particular
importance in supersymmetric theories.
It was a big success of Gibbons {\it et al.} to have been able to
show that the \KY tensor,
which had long been known for relativists as rather mysterious structure,
can be understood as an object generating a `nongeneric' supersymmetry,
i.e. supersymmetry appearing only in specific spacetimes.
The corresponding supercharge $Q_f$ generated by the \KY tensor
is a superinvariant rather than merely a constant of motion.
The \KY tensor here is a 2-form, $f_{\mu\nu}=f_{[\mu\nu]}$,
which satisfies the \PF equation \cite{PF}
\begin{equation}
        D_{(\mu}f_{\nu)\lambda}=0.
        \label{PFeqn}
\end{equation}
It is also worth noting that the square of a \KY tensor makes the
associated Killing tensor $K_{\mu\nu}$ as
\begin{equation}
        K_{\mu\nu}=f_{\mu\lambda}f_\nu{}^\lambda.
        \label{K=ff}
\end{equation}
It is of some interest that
so-called the Carter's constant $K_{\mu\nu}u^\mu u^\nu$ is the bosonic
sector of square of $Q_f$, $\brace{Q_f,Q_f}$.
($u^\mu$ is the particle's tangent.)
We may call 2-forms satisfying Eq.\reff{PFeqn} \KY tensors of usual type,
whereas we call $r$-forms satisfying similar equation
\begin{equation}
    D_{(\mu_1}f_{\mu_2)\mu_3\cdots\mu_{r+1}}=0
    \label{vr}
\end{equation}
{\it \KY tensors of valence} $r$ \cite{DR,Ya}.

In this paper we discuss the role of these generalized \KY tensors,
with the framework extended to include electromagnetic interactions.

We shall first retrace the argument in \cite{GRvH}
with the extended framework and see the
manifestation of electromagnetic interactions.
One notable consequence would be the condition of the electromagnetic
tensor $F_{\mu\nu}$ to maintain the nongeneric supersymmetry.
Using the \KY tensor, $f_{\mu\nu}$, this condition will be expressed as
\begin{equation}
    f^\lambda{}_{[\mu}F_{\nu]\lambda}=0.
    \label{i-1}
\end{equation}
This has also been known in the approach using the conformal
Killing spinor \cite{HPSW}, $\chi_{AB}$, as the condition to maintain
the constant of motion, $\chi=\chi_{AB}\lambda^A\lambda^B$, along
the null geodesics generated by $\lambda^A\bar{\lambda}^{A'}$.
In 2-spinor notation, this condition is expressed as
\begin{equation}
    \chi_{B(A}\phi_{C)}{}^B=0,
    \label{i-2}
\end{equation}
where $\phi_{CB}$ is the electromagnetic spinor.
If the conformal Killing spinor $\chi_{AB}$ satisfies
a subsidiary condition
$\nabla_{A'C}\chi^C{}_A-\nabla_{AC'}\bar{\chi}^{C'}{}_{A'}=0$,
then $\chi_{AB}$ is called the Killing spinor in strong sense \cite{CM}
and coincides with the spinor version of the \KY tensor of usual type.
With such $\chi_{AB}$, Eq.\reff{i-2} is equivalent to Eq.\reff{i-1}.
It is worth noticing that
the condition implies that the principal null directions (the PND)
of the electromagnetic field must be aligned with those of
the Killing spinor \cite{HPSW}.

We then discuss the role of
\KY tensors of valence $r$, $f_{\mu_1\cdots\mu_r}$.
We know that, in the usual relativistic point particle theory,
Killing tensors imply constants of motion, i.e., if the spacetime
admits a Killing tensor $K_{\mu_1\cdots\mu_r}$ of valence $r$,
then the phase space function
$K_{\mu_1\cdots\mu_r}u^{\mu_1}\cdots u^{\mu_r}$
is constant along the geodesic \cite{WP}.
What we point out in this paper is a counterpart to this
in the supersymmetric theory.
We will find the one-to-one correspondence between \KY tensors
and superinvariants, of which forms are rather nontrivial.
We also examine the brackets of such superinvariants with generic
constants of motion, and thereby discuss the associated constants
of motion with such superinvariants.

Although it has no obstacles in passing to quantum mechanics,
we shall concentrate on classical analysis.
Nevertheless, we know that \KY tensors can play a key role
in the Dirac's theory on a curved spacetime \cite{CM}.
Our results may strengthen the connection of \KY tensors with
the supersymmetric classical and quantum mechanics
on curved manifolds.

The plan of this paper is as follows.
In sect.\ref{sec2} we establish the canonical formulation
of pseudo-classical charged spinning particles in an arbitrary
background spacetime, using Grassmann-valued pseudo-Lorentz
vector to describe the spin degrees of freedom.
In sect.\ref{sec3} we formulate component equations for
extra symmetries, i.e. for constants of motion and superinvariants.
In sect.\ref{sec4} we see if the nongeneric supersymmetry
survives when the electromagnetic
interactions are taken into account.
Sections \ref{sec2} through \ref{sec4} are also reviews for the
treatment of the symmetries of supersymmetric point particle theory.
In sect.\ref{sec5} we establish the role of general \KY tensor
of valence $r$.
We in sect.\ref{sec:aYr} consider the possibility of spacetimes admitting
a \KY tensor to have larger symmetries.
Finally, sect.\ref{secConc} is devoted to conclusions.

\section{The pseudo-classical description of a
charged Dirac particle} \label{sec2}

In this section, we establish the pseudo-classical description of
our charged Dirac particle.
Note first that, while usual point particle is described by
its point $x^\mu$ on a Lorenzian manifold $(M,\dg\mu\nu)$,
our pseudo-classical (charged) Dirac particle has also a freedom
of spin which is represented by a Grassmann-valued pseudo-Lorentz
vector $\xi^a$.
Such descriptions was considered in Refs.\cite{BM,Casa,BCL,BDZVH,BDVH},
and in particular we shall employ the linearized Lagrangian
treated in Ref.\cite{BDVH}.
We thus start with the Lagrangian,
\begin{equation}
    L={m\over2}\dg\mu\nu\ux\mu\ux\nu+eA_\mu\ux\mu+
      {i\over2}\paren{\xi_a\frac{D\xi^a}{d\tau}-{e\over m}F_{ab}\xi^a\xi^b},
    \label{Lagrangian}
\end{equation}
where $m$ and $e$ are, respectively, the mass and the charge of
a particle, and $A_\mu(x)$ and $F_{\mu\nu}(x)$, respectively,
the vector potential and the field strength of the electromagnetic
field, both of which are considered as external fields, and so is the
spacetime metric $\dg\mu\nu(x)$.
Greek and Latin indices refer to world and Lorentz indices, respectively,
and are converted into each other by the vielbein $e_a{}^\mu$.
The dot over $x^\mu$ represents the derivative with respect to
a parameter $\tau$,
while $D\xi^a/d\tau$ represents the covariant derivative with respect to
$\tau$;
\begin{equation}
    \frac{D\xi^a}{d\tau}=\dot{\xi^a}+\omega^a{}_{b\mu}\xi^b\dot{x}^\mu,
\end{equation}
where $\omega_{ab\mu}$ is the connection 1-form.

Since our Lagrangian is a gauge-fixed one, we have to
add appropriate constraints.
One is given by
\begin{equation}
    H\equiv{m\over2}\dg\mu\nu\dot{x}^\mu\dot{x}^\nu+
    \frac{ie}{2m}F_{ab}\xi^a\xi^b\approx -{m\over2},
\end{equation}
which ensures the time-reparametrization invariance.
Also, this tells that the parameter $\tau$ is a generalization
of the proper time.
The other constraint is
\begin{equation}
    Q\equiv e_{a\mu}\dot{x}^\mu\xi^a\approx0,
        \label{Q1}
\end{equation}
which generates the supersymmetry transformation
\footnote{The full Lagrangian contains,
        apart from the Lagrange multipliers,
        a Grassmann-valued pseudo-Lorents scaler $\xi_5$ as a variable,
        which must be introduced to `carry' the mass.
        The supercharge \reff{Q1} should have been
        $Q\equiv e_{a\mu}\dot{x}^\mu\xi^a+\xi_5\approx0$
        to recover the massive Dirac equation when quantized.
        However, in the present gauge, $\xi_5$ is found to be a constant,
        so that the subsequent classical analysis will not be affected
        with $\xi_5$ suppressed.
        The constancy of $\xi_5$ will appear as the existence of
        the chiral charge (See Eq.\reff{Gammastar}).
}.
The equations of motion derived from the above Lagrangian will be invariant
under the transformation generated through
appropriate Poisson-Dirac bracket with the above constraints.
Variable $\xi^a$ is the superpartner of $x^\mu$ for the supersymmetry
transformation generated by $Q$.
Our Lagrangian gives, in conjunction with the constraints $H$ and $Q$,
the pseudo-classical description of charged Dirac (spinning) particles.

Since the conjugate momenta are
\begin{equation}
    p_\mu=\Del{L}{\dot{x}^\mu}=m\dg\mu\nu\dot{x}^\nu+\omega_\mu+eA_\mu,
    \quad
    \pi_a=\Del{L}{\dot{\xi}^a}=-\frac i2\xi_a
\label{mo}
\end{equation}
with $\omega_\mu\equiv(i/2)\omega_{ab\mu}\xi^a\xi^b$,
the second class constraint for $\pi_a$
yields the following Poisson-Dirac bracket;
\begin{equation}
  \brace{F,G}=\Del{F}{x^\mu}\Del{G}{p_\mu}-\Del{F}{p_\mu}\Del{G}{x^\mu}
  +i(\del F/\del \xi^a)\Del{G}{\xi_a},
  \label{PD1}
\end{equation}
where $(\del F/\del \xi^a)$ is a right differentiation which will take
the opposite sign to $\Del{F}{\xi^a}$ when $F$ is Grassmann-odd.
With this bracket, we can check the canonical relations,
$\brace{x^\mu,p_\nu}=\delta^\mu_\nu$
and $\brace{\xi^a,\xi^b}=-i\eta^{ab}$.
For convenience, we introduce
the gauge-covariant variable $\Pi_\mu$ defined by
\begin{equation}
  \Pi_\mu=p_\mu-\omega_\mu-eA_\mu(=m\dg\mu\nu\ux\nu).
  \label{defPi}
\end{equation}
With this variable, the bracket becomes
\begin{equation}
    \brace{F,G}=({\cal D}_\mu F)\Del{G}{\Pi_\mu}-
    \Del{F}{\Pi_\mu}({\cal D}_\mu G)+
    (R_{\mu\nu}+eF_{\mu\nu})\Del{F}{\Pi_\mu}\Del{G}{\Pi_\nu}+
    i({\del F}/{\del\xi^a})\Del{G}{\xi_a},
    \label{PDcov}
\end{equation}
where we have defined the spin-valued Riemann tensor
\begin{equation}
    R_{\mu\nu}\equiv{i\over2}R_{ab\mu\nu}\xi^a\xi^b
\end{equation}
and the phase space covariant derivative operator
\begin{equation}
    {\cal D}_\mu F\equiv\Del{F}{x^\mu}+
    \Pi_\lambda\Gamma^{\lambda}_{\mu\nu}\Del{F}{\Pi_\nu}-
    \omega^a{}_{b\mu}\xi^b\Del{F}{\xi^a}.
\end{equation}
Now, with this bracket it is easy to see for the constraints
\begin{equation}
    H=\rcp{2m}\ug\mu\nu\Pi_\mu\Pi_\nu+\frac{ie}{2m}F_{ab}\xi^a\xi^b
    \approx-\frac{m}{2}
    \label{H}
\end{equation}
and
\begin{equation}
    Q=\rcp{m}e_a{}^\mu\Pi_\mu\xi^a\approx0
    \label{Q}
\end{equation}
that the usual supersymmetry algebra
\begin{equation}
  \brace{Q,H}=0, \quad \brace{Q,Q}=-\frac{2i}{m}H.
  \label{susy-alge}
\end{equation}
holds.

\section{Generalized Killing equations and their square `roots'} \label{sec3}

In this section we write down the equations for constants of motion
and superinvariants,
which will be applied in the subsequent sections.

First, for any constant of motion $J(x,\Pi,\xi)$,
the bracket with $H$ vanishs, $\brace{H,J}=0$.
With the bracket \reff{PDcov}, this reduces to
\begin{equation}
  \Pi^\mu\brace{{\cal D}_\mu J-
    \Del{J}{\Pi_\nu}\paren{R_{\mu\nu}+eF_{\mu\nu}}}=
  eF_\mu\Del{J}{\Pi_\mu}+e\,\xi^aF_a{}^b\Del{J}{\xi^b},
\end{equation}
where $F_\mu\equiv(i/2)(D_\mu F_{ab})\xi^a\xi^b$.
Following \cite{GRvH}, let us expand $J(x,p,\Pi)$ in powers of $\Pi_\mu$;
\begin{equation}
  J=\sum^\infty_{n=0}\rcp{n!}J^{(n)\mu_1\cdots\mu_n}
  (x,\xi)\Pi_{\mu_1}\!\cdots\Pi_{\mu_n}.
  \label{Jpan}
\end{equation}
Then we have for the coefficients $J^{(n)\mu_1\cdots\mu_n}$
the following generalized Killing equations;
\begin{eqnarray}
  D_{(\mu}J^{(n)}_{\mu_1\cdots\mu_n)}&-&
  \omega^a{}_{b(\mu}\xi^b\Del{J^{(n)}_{\mu_1\cdots\mu_n)}}{\xi^a}
  \nonumber\\ &=&
  -(R_{\nu(\mu}+eF_{\nu(\mu})J^{(n+1)}_{\mu_1\cdots\mu_n)}{}^\nu+
  eF_\nu\rcp{n+1}J^{(n+2)}_{\mu_1\cdots\mu_n\mu}{}^\nu+
  e\,\xi^aF_a{}^b\rcp{n+1}\Del{J^{(n+1)}_{\mu_1\cdots\mu_n\mu}}{\xi^b}
  \label{gKilling}
\end{eqnarray}
and
\begin{equation}
  F_\mu J^{(1)\mu}+\xi^aF_a{}^b\Del{J^{(0)}}{\xi^b}=0.
  \label{gKilling2}
\end{equation}
These are a direct generalization of Eq.(41) in \cite{GRvH},
though there exist some differences of sign due to the difference of
sign convention of connection 1-form $\omega_{ab\mu}$.
We shall refer to Eq.\reff{gKilling2} as $n=-1$ component of the
generalized Killing equation \reff{gKilling}.

The equation for superinvariants is derived from the equation
$\brace{Q,J}=0$.
Such a superinvariant $J$ is automatically a constant of motion,
i.e., $\brace{H,J}=0$, as confirmed by the Jacobi identity
with Eq.\reff{susy-alge}.
Again, with the bracket \reff{PDcov}, we have
\begin{equation}
  \xi^\mu\paren{{\cal D}_\mu J-
    eF_{\mu\nu}\Del{J}{\Pi_\nu}}+i\Pi^a\Del{J}{\xi^a}=0.
  \label{sinv}
\end{equation}
Expanding $J^{(n)\mu_1\cdots\mu_n}$ in powers of $\xi^a$ and letting
the coefficients be $f^{(m,n)\mu_1\cdots\mu_n}_{a_1\cdots a_m}(x)$,
i.e.,
\begin{equation}
  J=\sum^\infty_{m,n=0}\frac{i^{[{m\over2}]}}{m!n!}\xi^{a_1}\cdots\xi^{a_m}
  f^{(m,n)\mu_1\cdots\mu_n}_{a_1\cdots a_m}
  (x)\Pi_{\mu_1}\!\cdots\Pi_{\mu_n},
  \label{Jpan2}
\end{equation}
we obtain from Eq.\reff{sinv} the component equation;
\begin{equation}
    me_{[a}{}^\mu D_\mu f^{(m-1,n)\mu_1\cdots\mu_n}_{a_1\cdots a_{m-1}]}-
    meF_{\mu\nu}e_{[a}{}^\mu
        f^{(m-1,n+1)\mu_1\cdots\mu_n\nu}_{a_1\cdots a_{m-1}]}-
    nf^{(m+1,n-1)(\mu_1\cdots\mu_{n-1}}_{baa_1\cdots a_{m-1}}
        e^{b\mu_n)}=0.
\label{sinv2}
\end{equation}
We may call this equation the generalized \PF equation.
This is also sometimes referred to as the square roots of the generalized
Killing equation \cite{GRvH}.

\section{Nongeneric supersymmetries} \label{sec4}

Following Ref.\cite{GRvH}, we search for nongeneric supersymmetry
with the generator of the form
\begin{equation}
    Q_f=\xi^af_a{}^\mu(x)\Pi_\mu+\frac{i}{3!}c_{abc}(x)\xi^a\xi^b\xi^c
    +h_a(x)\xi^a,
    \label{Q_f}
\end{equation}
where $f_a{}^\mu(x), c_{abc}(x)$ and $h_a(x)$ are functions of $x^\mu$.
The first term of the right side is an analogue of the
supercharge $Q$ (see \reff{Q}).
This charge generates the supersymmetry transformation such as
\begin{equation}
        \delta x^\mu=i\epsilon\brace{Q_f,x^\mu}=-i\epsilon\xi^a f_a{}^\mu,
        \label{susytrans}
\end{equation}
where the infinitesimal parameter $\epsilon$ of the transformation
is Grassmann-odd.

We do not investigate the conditions that $Q_f$ commute with $H$,
but with $Q$, since
we are interested in the \KY tensor, which will be found to have
close relationship with a superinvariant rather than a constant
of motion.

We evaluate all nontrivial components of Eq.\reff{sinv2}
with $J$ being $Q_f$ given by Eq.\reff{Q_f}.
First of all, component $(m,n)=(0,1)$ gives
$h_be^{b\mu}=0$. That is, $h_a$ must vanish.
Next, look at component $(m,n)=(0,2)$, giving
$f_b{}^{(\mu_1}e^{b\mu_2)}=0$.
Introducing $f_{\mu\nu}\equiv f_{a\mu}e^a{}_\nu$,
this implies that $f_{\mu\nu}$ must be antisymmetric,
\begin{equation}
  f_{(\mu\nu)}=0.
  \label{anti}
\end{equation}
Then, look at component $(m,n)=(2,1)$, which gives
\begin{equation}
  2e_{[a}{}^\mu D_\mu f_{b]}{}^{\mu_1}-c_{cab}e^{c\mu_1}=0.
  \label{tmp1}
\end{equation}
Again, it will be useful to introduce the world-indices version of
$c_{abc}$, $c_{\mu\nu\lambda}=c_{abc}e^a{}_\mu e^b{}_\nu e^c{}_\lambda$.
Then, we observe from Eq.\reff{tmp1} that $D_{[\mu}f_{\nu]\lambda}$
must be skew-symmetric in accordance with the skew-symmetry of
$c_{abc}$ or $c_{\mu\nu\lambda}$, so that taking Eq.\reff{anti}
into account we have the \PF equation \reff{PFeqn}.
This implies that $f_{\mu\nu}$
is the \KY tensor, as in the vacuum case.
With Eqs.\reff{PFeqn} and \reff{anti}, Eq.\reff{tmp1} yields
\begin{equation}
  c_{\mu\nu\lambda}=-2D_\mu f_{\nu\lambda}\; (=-2D_{[\mu}f_{\nu\lambda]}),
  \label{c}
\end{equation}
so that $c_{\mu\nu\lambda}$ is given by differentiation of $f_{\mu\nu}$
and is exact.
It is easy to see that component $(m,n)=(4,0)$, which is
$D_{[\mu}c_{\nu\lambda\sigma]}=0$,
becomes trivial, since $c_{\nu\lambda\sigma}$ is exact.
Finally, component $(m,n)=(2,0)$ gives Eq.\reff{i-1},
the condition for the coincidence of PND's alignment
of the electromagnetic spinor and the Killing spinor.

After all, we have the final form of $Q_f$;
\begin{equation}
    Q_f=\xi^\nu f^\mu{}_\nu\Pi_\mu
        -\frac{i}{3}\xi^\mu\xi^\nu\xi^\lambda D_{[\mu}f_{\nu\lambda]}.
        \label{Qf2}
\end{equation}
Difference from the vacuum case is only the
definition \reff{defPi} of $\Pi_\mu$ in terms of $p_\mu$,
if Eq.\reff{i-1} holds.
This type of superinvariants exists
in the \KN spacetime \cite{GRvH,CM}.
Although there are not so many physically interpretable
spacetimes which admit a \KY
tensor \cite{DR,MT}, another such interesting example would be
the Taub-NUT spacetime \cite{Vis,vH},
which admits four independent \KY tensors \cite{vH}.

It is straightforward to calculate the constant of motion
$K\equiv\frac{i}{2}\brace{Q_f,Q_f}$,
which is given by
\begin{eqnarray}
  K\wa\rcp2 f^\mu{}_\lambda f^{\nu\lambda}\Pi_\mu\Pi_\nu \nonumber \\
  &+&\frac{i}{2}\xi^\mu\xi^\nu\brace{2(D_\lambda f^\sigma{}_\mu)
    f^{\lambda}{}_{\nu}\Pi_\sigma
    +c_{\mu\nu\lambda}f^{\sigma\lambda}\Pi_\sigma
    +eF_{\lambda\sigma}f^\lambda{}_\mu f^\sigma{}_\nu} \nonumber \\
  &+&\rcp4\xi^\mu\xi^\nu\xi^\lambda\xi^\sigma
  \brace{R_{\mu\nu\kappa\omega}f^\kappa{}_\lambda f^\omega{}_\sigma
    -\rcp2 c_{\mu\nu\kappa}c_{\lambda\sigma}{}^\kappa},
  \label{K}
\end{eqnarray}
where we have used the relation
$D_\mu c_{abc}=3f_{[c}{}^\nu R_{ab]\mu\nu}$.
As expected, the bosonic sector of $K$ is the
quadratic $\rcp2 K_{\mu\nu}u^\mu u^\nu$ (with Eq.\reff{K=ff}).

\section{Generalized Killing-Yano tensors and corresponding superinvariants}
\label{sec5}

We are now in a position to discuss the role of general \KY tensors.
This will respond to the question of
how profound the connection of the appearance
of nongeneric supersymmetries and the existence of the \KY tensors is,
and will give a useful tool in investigating a supersymmetric dynamical system.

What we want to note first is
the \KY tensor of valence $d=\dim(M)$,
which is generic and coincides with the volume form
$\epsilon_{\mu_1\cdots\mu_d}$ up to a constant factor.
This object appears in two generic constants of motion,
the chiral charge \cite{RvH}
\begin{equation}
    \Gamma_*\equiv -\frac{i^{[{d\over2}]}}{d!}
        \epsilon_{a_1\cdots a_d}\xi^{a_1}\cdots\xi^{a_d},
    \label{Gammastar}
\end{equation}
and the dual supercharge
\begin{equation}
    Q^*=i\brace{Q,\Gamma_*}=
    \frac{-i^{[{d\over2}]}}{(d-1)!}
        \epsilon_{a_1\cdots a_d}e^{a_1\mu}\Pi_\mu\xi^{a_2}\cdots\xi^{a_d}.
    \label{Qstar}
\end{equation}
(The form of these generic charges is regardless of the existence of
electromagnetic interactions.)
We note that the dual supercharge $\dualQ$ is superinvariant, and
the form of it, Eq.\reff{Qstar}, is similar to
that of the nongeneric supercharge \reff{Qf2}.

An analogy leads us to try to find supercharges in the form
\begin{equation}
        J=\frac{i^{[\frac{r-1}{2}]}}{(r-1)!}
        f_{a_1\cdots a_{r-1}}^{(r-1,1)}{}^\mu\Pi_\mu
        \xi^{a_1}\cdots\xi^{a_{r-1}}
        +\frac{i^{[\frac{r+1}{2}]}}{(r+1)!}
        f_{a_1\cdots a_{r+1}}^{(r+1,1)}\xi^{a_1}\cdots\xi^{a_{r+1}}.
        \label{Yr1}
\end{equation}
It is easy to examine Eq.\reff{sinv2} for Eq.\reff{Yr1}.
Calculations done are completely parallel to those in the
previous section.
The following theorem summerizes the result.

\medskip
\noindent {\bf Theorem: }
{\it
If the spacetime admits a \KY tensor of valence $r$, $f_{\mu_1\cdots\mu_r}$,
and the electromagnetic field $F_{\mu\nu}$ satisfies the condition
\begin{equation}
        F_{\nu[\mu_r}f_{\mu_1\cdots\mu_{r-1}]}{}^\nu=0,
        \label{Ff}
\end{equation}
then the function
\begin{equation}
        Y_r=\xi^{\mu_2}\cdots\xi^{\mu_r}f_{\mu_1\cdots\mu_r}\Pi^{\mu_1}
                -\frac{i}{r+1}\xi^{\mu_1}\cdots\xi^{\mu_{r+1}}
                D_{[\mu_1}f_{\mu_2\cdots\mu_{r+1}]} \label{Yr}
\end{equation}
is a superinvariant, $\brace{Q,Y_r}=0$,
for the bracket defined by Eq.\reff{PDcov}.
The converse also holds.
}
\medskip

\noindent
Here, $f_{\mu_1\cdots\mu_r}$ corresponds to
$f_{a_1\cdots a_{r-1}}^{(r-1,1)}{}^\nu
g_{\nu\mu_1}e^{a_1}{}_{\mu_2}\cdots e^{a_{r-1}}{}_{\mu_r}$.

Thus, we know
\begin{equation}
        \dualQ= \frac{-i^{[{d\over2}]}}{(d-1)!}Y_d,
        \label{Yd}
\end{equation}
provided that $f_{\mu_1\cdots\mu_d}=\epsilon_{\mu_1\cdots\mu_d}$.

Eq.\reff{vr} implies that a \KY tensor of valence 1
is a usual Killing vector.
Let $\zeta^\mu$ be a Killing vector, then
it is a direct consequence of the theorem that
\begin{equation}
        Y_1=\zeta^\mu\Pi_\mu
        -\frac i2\xi^\mu\xi^\nu D_\mu\zeta_\nu
        \label{Qzeta}
\end{equation}
is superinvariant, if
\begin{equation}
        F_{\mu\nu}\zeta^\nu=0
        \label{cc}
\end{equation}
holds.
However, if we consider an alternative function
\begin{equation}
        J_\zeta=\zeta^\mu(\Pi_\mu+eA_\mu)
        -\frac i2\xi^\mu\xi^\nu D_\mu\zeta_\nu,
        \label{Jzeta2}
\end{equation}
this is superinvariant, regardless of Eq.\reff{cc}, provided
that the Lie derivative of the vector potential
with respect to $\zeta^\mu$ vanishes, $\Lie\zeta A_\mu=0$.
We would have got Eq.\reff{Jzeta2} as a result of trying
to obtain a constant of motion (not superinvariant) associated
with a Killing vector, using Eq.\reff{gKilling}
(cf. Ref.\cite{HPSW}), however Eq.\reff{Jzeta2} happens to be
superinvariant.
This is a special feature for $r=1$.

\def\Yr{\mbox{$Y_r$}}
\section{The constants of motion associated with \Yr}
\label{sec:aYr}

As already established, if a spacetime admits a \KY tensor of valence r
and if the electromagnetic tensor satisfies Eq.\reff{Ff},
then $Y_r$ is a constant of motion of a spinning particle in the spacetime.
Possibly, there exist other constants of motion associated with $Y_r$, i.e.,
there may exist nonvanishing brackets of $Y_r$ with other known
constants of motion.
We here discuss generic feature of such constants of motion, i.e.,
we suppose there are no nongeneric constants of motion other than
$Y_r$ for specific value of $r$.

It is obvious that we can construct such constants of motion first by
taking brackets of $Y_r$ with the four generic constants of motion,
$H$, $Q$, $\chiral$, and $\dualQ$.
Since $Y_r$ is a (super)invariant, we cannot use $H$ and $Q$ for the
present purpose.
Moreover, since $\dualQ$ has connection with $\chiral$ through
$\dualQ=i\brace{Q,\chiral}$, we do not have to discuss
$\dualQ$ and $\chiral$ separately.
In fact, if $\dim(M)=d$, we have
\begin{eqnarray}
        \brace{Y_r,\dualQ}\wa i\brace{Y_r,\brace{Q,\chiral}} \nonumber \\
        \wa -i(-1)^d\brace{\chiral,\brace{Y_r,Q}}
        -i(-1)^{d(r-1)}\brace{Q,\brace{\chiral,Y_r}} \nonumber \\
        \wa i\brace{Q,\brace{Y_r,\chiral}} \nonumber \\
        (\wa \brace{Q,Y_r^*}), \label{duality}
\end{eqnarray}
where the second line is the consequence of the Jacobi identity
and we have defined the dual of $Y_r$ as
$Y_r^*\equiv i\brace{Y_r,\chiral}$.
Hence the bracket of $Y_r$ with the dual supercharge $\dualQ$ is also the
bracket of the supercharge $Q$ and the dual of $Y_r$, so that
we only need to start with seeing if there exist nonvanishing duals of $Y_r$.

Since ${\cal D}_\mu\chiral=0$ and $(\partial\chiral/\partial{\Pi_\mu})=0$,
we have
\begin{equation}
        Y_r^*=(-1)^r\Del{Y_r}{\xi^a}\Del{\chiral}{\xi_a}.
        \label{Yrstar}
\end{equation}
The numbers of the Grassmann vectors in $Y_r$ and $\chiral$ are,
respectively, $r-1$ (the least number) and $d$, so that
that of Eq.\reff{Yrstar} is $r-1+d-2=r+d-3$,
which must be not greater than $d$ in order that $Y_r^*$ do not vanish.
We can therefore have nonvanishing $Y_r^*$ only for $r\leq3$.
However, for $r=1,3$, we find by direct calculations that
$Y_r^*$ vanishes after all.
Thus, we can generate new constants of motion only for $Y_2=Q_f$.

We can immediately calculate $Y_2^*=Q_f^*$, which gives
\begin{equation}
        Q_f^*={-i^{[\frac d2]} \over (d-1)!} \epsilon_{a_1\cdots a_d}
        e^{a_1\mu}f^\nu{}_\mu\Pi_\nu \xi^{a_2}\cdots \xi^{a_d}.
        \label{Qfs}
\end{equation}
Then we can calculate the bracket of $Q_f^*$ with the supercharge,
for which we define $\Af$;
\begin{eqnarray}
        \Af \wb m\brace{Q,Q_f^*}=m\brace{Q_f,\dualQ} \nonumber \\
        \wa {eF_{\mu\nu}f^{\nu\mu}} \chiral
        -{i^{[{d\over2}]+1} \over (d-2)!} \epsilon_{a_1\cdots a_d}
        e^{a_1\mu}\Pi_\mu e^{a_2\nu}f^\lambda{}_\nu\Pi_\lambda
        \xi^{a_3}\cdots\xi^{a_d}.
        \label{af}
\end{eqnarray}
This is the end of our construction --- $\chiral$, $Q$, $\dualQ$,
$Q_f$, $Q_f^*$ and $\Af$ with $K$ and $H$ constitute a
closed algebra $\alg$, where $K$ is defined in Eq.\reff{K}
and $H$ is the Hamiltonian \reff{H}.
Fig.1 summerizes the relation in $\alg$.

Of particular interest is the maximal abelian subalgebra, $\salg$,
of $\alg$.
We can easily find that $\chiral$, $\dualQ$,
$Q_f^*$ and $\Af$ with $K$ and $H$ constitute
such an algebra $\salg$ and the dimension of it is six.
In the \KN spacetime, we have another two commuting constants of motion,
$J_\zeta$ and $J_\psi$ coming from
the two commuting Killing vectors $\zeta$ and $\psi$,
where, say, $\zeta$ is timelike and $\psi$ is the spacelike Killing
vector generating closed orbits.
Functions $J_\zeta$ and $J_\psi$ also commute with all elements of $\salg$,
and with $\salg$ form the largest abelian algebra.
This is a classical justification of the separability of
the Dirac equation in the \KN spacetime \cite{Cha,PT}.
We can easily find that the Taub-NUT spacetime is also
in the same situation.

\begin{figure}[tbp]
  \begin{center}
    \leavevmode
\epsfysize=8cm
\epsfbox{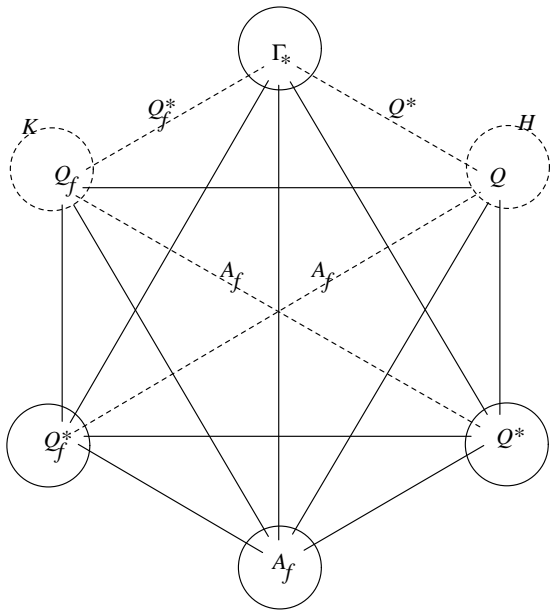}
  \end{center}
  \caption{The algebra $\alg$. Real lines stand for vanishing of the \PD
brackets, whereas dashed lines for non-zeros. Characters above dashed lines
are reminders of the non-vanishing brackets, e.g., the bracket of $\chiral$
and $Q$ is proportional to $\dualQ$. Real circles stand for vanishing
of the \PD brackets with oneself, whereas dashed circles for non-zeros.
The meaning of characters above the dashed circles is the same as
for the lines. $H$ and $K$ commute with any functions listed in the figure.
Note that functions $\chiral$, $\dualQ$, $\dualQ_f$, and $\Af$ are
mutually connected with real lines, so constitutes the (maximal)
abelian subalgebra of $\alg$. (See the last paragraph of this section.)}
  \label{fig:1}
\end{figure}

\section{Conclusions} \label{secConc}

We have shown that, if a spacetime admits a \KY tensor of valence $r$,
the spinning particles moving on it possess the superinvariant $Y_r$
defined by \reff{Yr}.
To hold the symmetry associated with the \KY tensor,
the electromagnetic tensor must satisfy Eq.\reff{Ff}.
The function $Y_2$, which is made from the \KY tensor of usual type,
is in a particular position, since only this can have nonvanishing
bracket with the chiral charge, $\chiral$, for which we can find the
associated other constants of motion, \reff{Qfs} and \reff{af}.

It should be noted that these facts do not depend on
the dimension of spacetime.
This enables us to apply our results to other supersymmetric systems,
e.g. supersymmetric cosmologies \cite{sucosmo},
where point particles in spacetimes
are replaced by points in the minisuperspaces.
Since the Lagrangians used there are not the same as the one used here,
the form of Eq.\reff{Yr} will vary.
However, it is plausible
that the \KY tensor can be a useful tool in investigating such systems.

Passing to quantum mechanics and the applications to supersymmetric
cosmologies will be discussed elsewhere.

\section*{Acknowledgments}
The author thanks Professor A. Hosoya for reading the manuscript
and for continuous encouragements.

\vfill

\end{document}